\documentclass[useAMS,usenatbib,usegraphicx]{mn2e}

\title[Capture and escape in the ERTBP]
{Capture and escape in the elliptic restricted three-body problem}
\author[S.~A.~Astakhov and D.~Farrelly]{Sergey A.~Astakhov$^{1,2}$\thanks{E-mail:
s.astakhov@fz-juelich.de; www.astakhov.newmail.ru} and David Farrelly$^{2}$\thanks{E-mail: david@habanero.chem.usu.edu}
\\
$^{1}$John von Neumann Institute for Computing, Forschungszentrum J\"ulich, D-52425 J\"ulich, Germany\\
$^{2}$Department of Chemistry and Biochemistry, Utah State University, UT 84322-0300, USA}

\begin{document}

\date{14 August 2004}

\maketitle

\begin{abstract}
  Several families of irregular moons orbit the giant planets. These
  moons are thought to have been captured into planetocentric orbits
  after straying into a region in which the planet's gravitation
  dominates solar perturbations (the Hill sphere). This mechanism
  requires a source of dissipation, such as gas-drag, in order to make
  capture permanent. However, capture by gas-drag requires that
  particles remain inside the Hill sphere long enough for dissipation
  to be effective. Recently we have proposed that in the circular
  restricted three-body problem particles may become caught up in
  `sticky' chaotic layers which tends to prolong their sojourn
  within the planet's Hill sphere thereby assisting capture.  Here we
  show that this mechanism survives perturbations due to the
  ellipticity of the planet's orbit. However, Monte Carlo simulations
  indicate that the planet's ability to capture moons decreases with
  increasing orbital eccentricity. At the actual Jupiter's orbital eccentricity,
  this effects in approximately an order of magnitude lower capture probability
  than estimated in the circular model. Eccentricities of planetary orbits
  in the Solar System are moderate but this is not necessarily the
  case for extrasolar planets which typically have rather eccentric
  orbits. Therefore, our findings suggest that these extrasolar
  planets are unlikely to have substantial populations of irregular
  moons.
\end{abstract}

\section{Introduction}

    While the large regular moons of the giant planets follow
    almost circular, low inclination, prograde orbits, the aptly named
    irregular moons tend to do the opposite: i.e., they often have
    highly eccentric, high inclination orbits which may be retrograde
    {\it or} prograde. These moons, therefore, provide actual
    observational examples of the complexities of three-body dynamics.
    Irregular moons are thought to have been captured during the early
    stages of the Solar System but the detailed mechanism has not been
    fully elucidated.  Based on a study of the circular restricted
    three-body problem we have recently proposed that chaos
    played a key role in the initial stages of capture \citep{cac2003}; in this
    mechanism -- called chaos assisted capture (CAC) -- particles may initially become entangled in chaotic
    layers which separate directly scattering from regular (bound)
    regions of phase space.  This temporary trapping serves to extend
    the lifetimes of particles within the Hill sphere thereby
    providing the breathing space necessary for relatively weak
    dissipative forces (e.g., gas-drag) to effect permanent capture.
    While this basic scenario may be responsible for the {\it initial}
    formation of families of irregular moons, other factors
    may affect their subsequent long term evolution including
    collisions; gravitational perturbations from other planets; and
    deviations from circularity of the planet's orbit. Here we
    investigate the effect of orbital ellipticity of the planet on the
    capture mechanism using Monte Carlo simulations and phase space
    visualisations based on the Fast Lyapunov Indicator. Two
    representative star-planet binaries are considered: the actual
    Sun-Jupiter system and a hypothetical extrasolar cousin of higher
    orbital eccentricity. We find that the chaos-assisted capture mechanism is robust to
    moderate ellipticities.

Recent discoveries of numerous irregular moons at giant planets
\citep{gladman2001,jewitt2003, kavel} have spurred the development
of theories of their origin, which is generally thought to involve
capture (\citealt{carruba2002,yokoyama2003,cac2003,burns2004};
\citealt*{nesvorny2004}; \citealt{neto2004, kavel}). These moons are
considered to be primordial pieces of the Solar System captured in
the amber of time.  While there is, as yet, no consensus on the
detailed capture mechanism of these minor bodies a number of recent
studies have, nevertheless, attemped to simulate their post-capture
evolution in order to explain salient aspects of their orbits. For
example, several attempts have recently been made to explain the
observed clustering of irregular satellites as being the result
either of the breakup of a larger parent body or from catastrophic
collisions between planetesimals \citep{nesvorny2003,
burns2004,nesvorny2004, kavel}. These hypotheses are supported
empirically by photometric surveys \citep{grav2004} which indicate
that members of the clusters have similar surface colours suggesting
that they may have had common progenitors. However, whatever
post-capture orbital evolution occurs {\it within} the Hill sphere
there has to have been an initial capture mechanism efficient enough
to populate the Hill sphere with moonlets in the first place. We
have recently proposed one such mechanism -- chaos-assisted capture
\citep{cac2003} -- in which long-term, but temporary, trapping
occurs in the `sticky' chaotic layers lying close to
Kolmogorov-Arnold-Moser \citep[KAM,][]{ll} tori embedded within the
planet's Hill sphere. If a particle gets entangled in one of these
layers then even a moderate level of dissipation (or, alternatively,
slow planetary growth, \citealt{neto2004}) may be sufficient to make
capture permanent.  It is likely that Comet-Shoemaker-Levy 9 was a
recent example of an object trapped in such a chaotic zone for the
best part of the last century \citep{chodas}. However, the absence
of a large gas cloud at Jupiter makes contemporary permanent capture
of such objects by gas-drag unfeasible.

Qualitatively, in the CAC model the capture probability of a planet
is largely determined by the volume of phase space (at a given
energy) taken up by chaotic zones within the Hill sphere.
Simulations in the circular restricted three-body problem (CRTBP)
show that capture becomes less probable at higher energies because
most of the phase space is directly scattering and relatively few
KAM tori survive; those that do are surrounded by relatively thin
chaotic layers which are less effective at trapping intruders
\citep{cac2003}. Factors that are not included in the CRTBP may also
affect the capture mechanism, e.g., the eccentricity of the parent
planet's orbit. A more realistic extension of CAC beyond the CRTBP
is, therefore, needed. As a first step we extend the model to the
{\it elliptic restricted three-body problem} (ERTBP, \citealt{sze,
llibre, benest}; \citealt{dvorak}; \citealt{scheeres};
\citealt{mako}) which allows us to model perturbations induced by
deviations of the planet's orbit from circularity. While orbital
eccentricities are relatively small, although not negligible, for
the giant planets of the Solar System they may be appreciable for
extrasolar planets (\citealt{schneider, marcy}; \citealt*{fischer};
\citealt{marzari, goldreich, beauge, michtchenko}). For example,
compare
 Jupiter's mean orbital eccentricity \footnote{\tt http://nssdc.gsfc.nasa.gov/planetary/planetfact.html}
$e=0.04839$ to typical extrasolar planet
eccentricities which lie in the range $e \sim 0.2-0.6$ \citep{tremaine}.
Up to now
there have been more than 100 extrasolar planetary systems detected
\footnote{\tt http://www.obspm.fr/planets \\ http://exoplanets.org} and an intriguing question is whether these massive
planets might harbour moons \citep{barnes, death, williams} which might
be habitable \citep*{habitable}. As an exemplary extrasolar captor, we
study a system similar in mass ratio to the idealized Sun-Jupiter binary but
which follows a highly elliptic orbit. This provides a comparative
estimate for the probability of capture and also suggests the ranges of energy and
orbital inclination over which extrasolar irregular moons might be
expected to exist.

The paper is organized as follows: Section 2 introduces the Hamiltonian
for the ERTBP which, in the limit of zero ellipticity, reduces to the
CRTBP. We work in a coordinate system whose origin coincides with the
planet and, in actual integrations, regularise the dynamics to deal
with two-body collisions \citep{Aarseth}. Because the ERTBP is explicitly time
dependent it is not possible to compute conventional Poincar\'e
surfaces of section \citep[SOS, ][]{ll} even in the planar limit so as to visualise
the structure of phase space. Therefore we use the notion of a Fast
Lyapunov Indicator \citep*[FLI, ][]{fro2000} to visualise the structure of phase space. Section
3 briefly discusses the CAC mechanism as applied to the CRTBP in
order to facilitate comparison with the ERTBP. In particular, FLI on the
surfaces of section are
computed which can be compared directly with SOS in the CRTBP. This is
done in Sec.~4 where Monte Carlo simulations of capture are
performed. Unlike in \citet{cac2003} in these simulations we do not
include dissipation and, instead, focus on the distributions of
particles that can be temporarily trapped in chaotic zones for very
long time periods. This avoids complications associated with the best
choice of dissipative force \citep[see, e.g., ][]{burns2004}. Finally,
conclusions are contained in Sec.~5.

\section{Hamiltonian and methods}
The CRTBP describes the dynamics of a test particle having
infinitesimal mass and moving in the gravitational field of two
massive bodies (the `primaries' -- e.g., a planet and a star) which
revolve around their center of mass on a circular orbit. The equations
of motion are, therefore, most naturally presented in a non-inertial
coordinate system that rotates with the mean motion of the
primaries \citep{murray}. In the rotating coordinate system the
positions of the primaries are fixed.  When the planet's orbit is
elliptic rather than circular a nonuniformly rotating-pulsating
coordinate system is commonly used. These new coordinates have the
felicitous property that, again, the positions of the primaries are
fixed; however the Hamiltonian is explicitly time-dependent
\citep{sze}.
\subsection{Hamiltonian}
\begin{figure}
\begin{center}
\includegraphics[width=6cm]{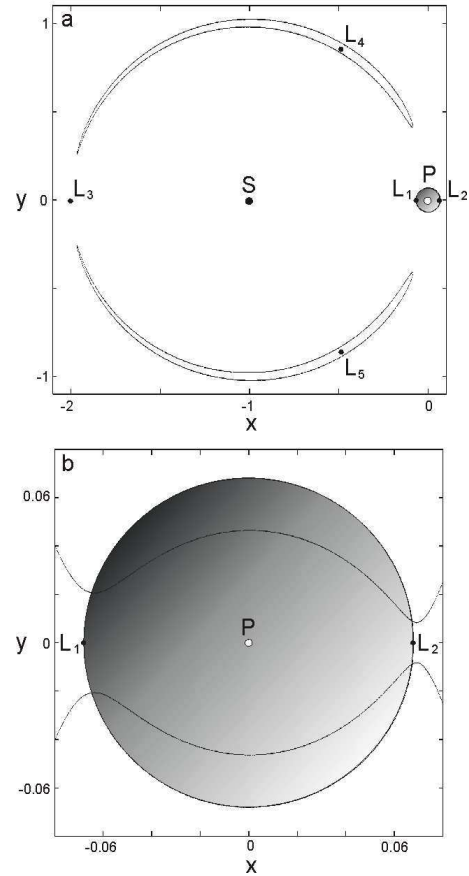}
\caption{\label{zvs} Level curves of the zero velocity surface at
high (a) and low (b) energies in the planetocentric planar CRTBP
along with the five Lagrange equillibrium points (labelled $L_1$ --
$L_5$). The circle shaded grey is the Hill sphere centered on the
planet (P). The star (S) is at $(-1,0)$.}
\end{center}
\end{figure}

For our purposes it is most convenient to locate the
origin at the planet (Fig.~\ref{zvs}) because angular momentum will be measured with
respect to the planet. Then, following \citet{sze} and \citet{llibre}, we obtain the planetocentric
ERTBP Hamiltonian $H_e$ after introducing an isotropically pulsating length scale

\[
H_e=E_e=\frac{1}{2}((p_x+y)^2 + (p_y-x)^2 + {p_z}^2 + z^2)-
\]
\[
\biggl(\frac{1-\mu }{\sqrt{(1+x)^2+y^2+z^2}} + \frac{\mu }{\sqrt{x^2+y^2+z^2}} +
\]
\begin{equation}
  \frac{1}{2}(x^2 + y^2 + z^2) +
   (1-\mu)x +\frac{1}{2}(1-\mu) \biggr)\bigg/(1+e\,\cos\,f). \label{ertbp}
\end{equation}

\noindent The semimajor axis of the orbit of the primaries $a_p$ has been scaled to unity and $e$
is the eccentricity of the planet's orbit; $\mu = m_{1}/(m_{1}+m_{2})$, where
$\mu, m_{1}$ and $m_{2}$ are the reduced mass and masses of the planet and
star, respectively ($\mu = 9.5359 \times 10^{-4}$ for Sun-Jupiter). The
true anomaly $f$, i.e. the planet's angular position measured from the
pericenter, is related to the physical time $t$ through \citep{sze}
\begin{equation}
\frac{df}{dt} = \frac{(1+e\,\cos\,f)^{2}}{(1-e^{2})^{3/2}}. \label{true}
\end{equation}

The (planet-centered) CRTBP Hamiltonian \citep{cac2003}

\[
H_c=E_c=\frac{1}{2}({p_x}^2 + {p_y}^2 + {p_z}^2)-(x\,p_{y}-y\,p_{x})-
\]
\[
\frac{1-\mu }{\sqrt{(1+x)^{2}+y^{2}+z^{2}}}-\frac{\mu }{\sqrt{x^{2}+y^{2}+z^{2}}}-
\]
\begin{equation}
 (1-\mu)x - \frac{1}{2}(1-\mu), \label{crtbp}
\end{equation}

\noindent is recovered when $e=0$. In both cases (\ref{ertbp}) and (\ref{crtbp}), $h_z = x\,p_{y}-y\,p_{x}$ is
the $z$-component of angular momentum ${\bf h}=(h_x,h_y,h_z)$ with
respect to the planet. Note that the orbital inclination $i = \arccos
{h_z}/{|{\bf h}|}$ is invariant under isotropic pulsating rescaling from CRTBP to ERTBP.
The orbit is said prograde if $i < \pi/2$ ($h_z >0$) and retrograde otherwise.

The Hamiltonian of the elliptic problem (\ref{ertbp}) is a periodic
function of the true anomaly $f$ (which plays the role of time) and,
hence, generates a non-autonomous dynamical system. Unlike the
circular problem, the ERTBP does not possess an energy integral and,
evidently, no such useful guiding concept as a static zero-velocity
surface \citep{murray} can be introduced. Furthermore, due to the extra
dimension associated with the explicit time dependence, construction of the
SOS and, indeed, any visual analysis
of phase space seems impossible even in the planar limit.

\subsection{Numerical methods}

We have performed numerical integrations in which fluxes of
particles are simulated as having passed from heliocentric orbits
into the region surrounding the planet as defined by the Hill
sphere. This region roughly corresponds to the region between the
points labelled $L_1$ and $L_2$ in Fig.~\ref{zvs}. In practice it is
convenient to go to extended phase space by introducing an
additional pair of conjugate variables -- `coordinate' $f$ and
`momentum' $p_f = -H$ \citep{ll} for which the new Hamiltonian
becomes conservative.  In our numerical simulations we have used
this method in combination with Levi-Civita (in 2D) and
Kustaanheimo-Stiefel (in 3D) regularising techniques
\citep{KS,Aarseth} to avoid problems associated with two-body
collisions \citep{nagler}. Numerical integrations were done using a
Bulirsch-Stoer adaptive integrator \citep{NR}.

\begin{figure*}
\begin{center}
\includegraphics[width=170mm]{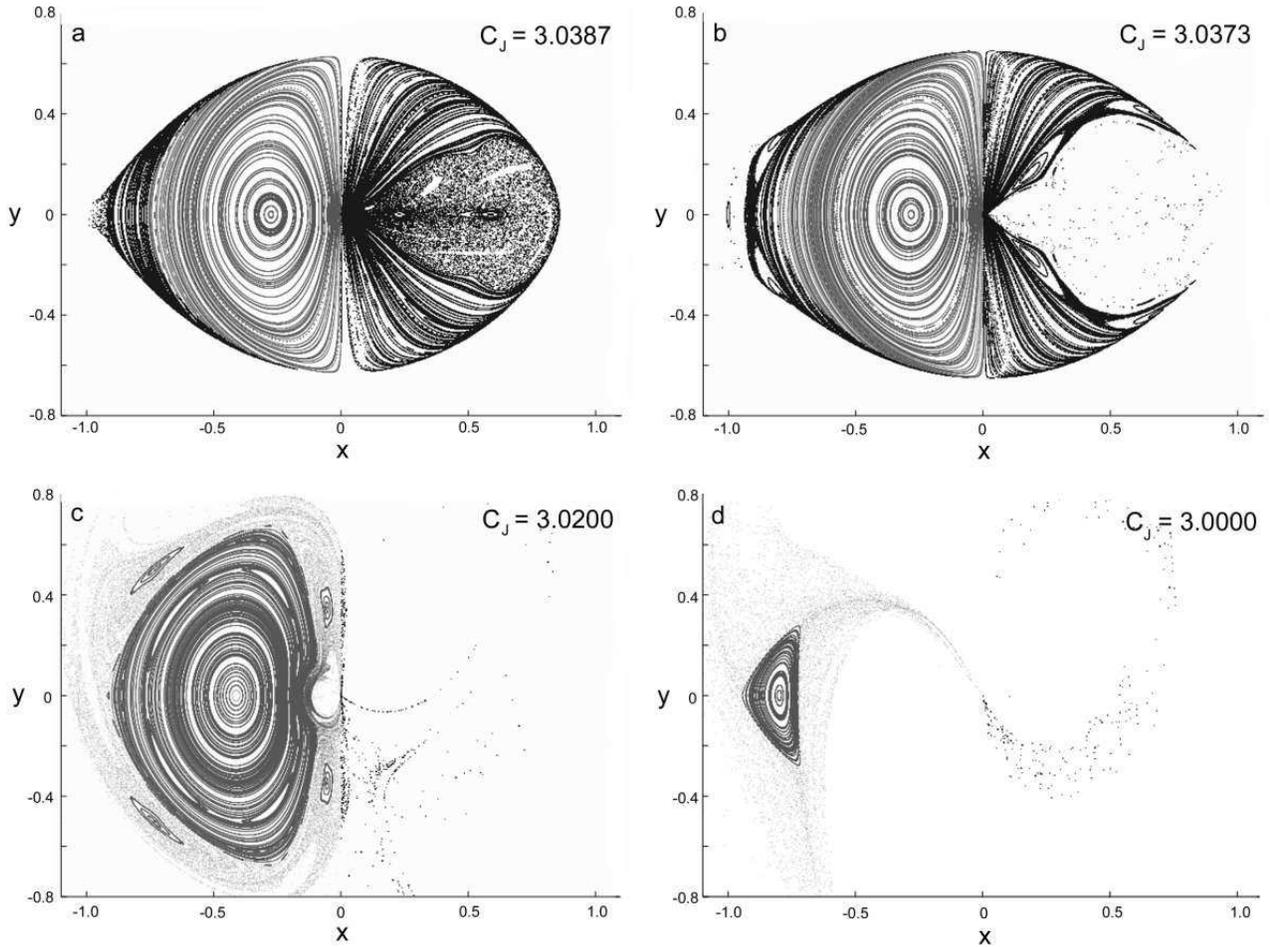}
\caption{\label{sos} Poincar\'e surfaces of section in the planar
CRTPB at four energies (Jacobi constants $C_J=-2E_c$). Initial
conditions were chosen randomly inside the Hill radius (here $R_H$
is scaled to 1) and integrated with the cut-off time $T_{cut}=36000$
years or until trajectories escaped. The SOS is the $x-y$ plane with
${p_x}=0, \dot{y}>0$. Points on the surface are coloured according
to the sign of angular momentum $h_z$ as the trajectories penetrate
the $x-y$ plane (black, prograde with $h_z > 0$; grey, retrograde
with $h_z < 0$).}
\end{center}
\end{figure*}

One approach to visualising phase space structures in systems with
greater than 2 degrees-of-freedom relies on computations of
short-time Lyapunov exponents (LE) over sets of initial conditions
of interest (for recent developments and relevant applications see,
e.g.,  \citet{fro2000}; \citet{sandor, dvorak}; \citet*{simo2003};
\citet{sandor2004} and references therein, although the idea of
using finite-time LE itself dates back at least to
\citealt{lorenz}). As was demonstrated by \citet{fro2000} various
dynamical regimes (including resonances) can be distinguished by
monitoring time profiles of a quantity called the Fast Lyapunov
Indicator. Given an $n$-dimensional continuous-time dynamical
system,
\begin{equation}
{d{\bf x}}/{dt} = {\bf F}({\bf x}, t), {\bf x} = (x_1, x_2,..., x_n), \label{EOM}
\end{equation}
the Fast Lyapunov Indicator is defined as \citep{fro2000}
\begin{equation}
FLI({\bf x}(0), {\bf v}(0), t) = \ln |{\bf v}(t)|, \label{fli}
\end{equation}
where {\bf v}(t) is a solution of the system of variational equations \citep*{tancredi}
\begin{equation}
\frac{d{\bf v}}{dt} = \biggl(\frac{\partial{\bf F}}{\partial{\bf x}} \biggr) {\bf v}. \label{variation}
\end{equation}

The complementary system (\ref{variation}) contains
spatial derivatives which only aggravate the singularities in the restricted three body problem (RTBP) and so the
use of regularisation becomes even more important. For this reason,
all calculations of FLI reported herein were made using regularised
versions of (\ref{EOM}) and (\ref{variation}).

As distinct from the largest Lyapunov characteristic exponent \citep{ll}
\begin{equation}
\lambda = \lim_{t\to\infty} \frac{1}{t} \ln \frac{|{\bf v}(t)|}{|{\bf v}(0)|}, \label{lyap}
\end{equation}
which just tends to zero for any regular orbit, the more sensitive FLI
can help discriminate between resonant and non-resonant regular orbits
\citep{fro2000}. Although, in practice, detection of resonances may be
tricky, since real differences in FLI for quasiperiodic and periodic
orbits of the same island are not always huge.

\section{Chaos assisted capture}

In this section we briefly summarize the CAC mechanism described in
\citet{cac2003} as applied to the CRTBP. Even though the basic
mechanism applies in three-dimensions, it is simpler here to outline
the general scheme in terms of the planar version of the CRTBP. Fig.~\ref{zvs}
shows the relevant zero velocity surface \citep[ZVS,][]{murray} together
with the five Lagrange equilibrium points.

\begin{figure}
\begin{center}
\includegraphics[width=7cm]{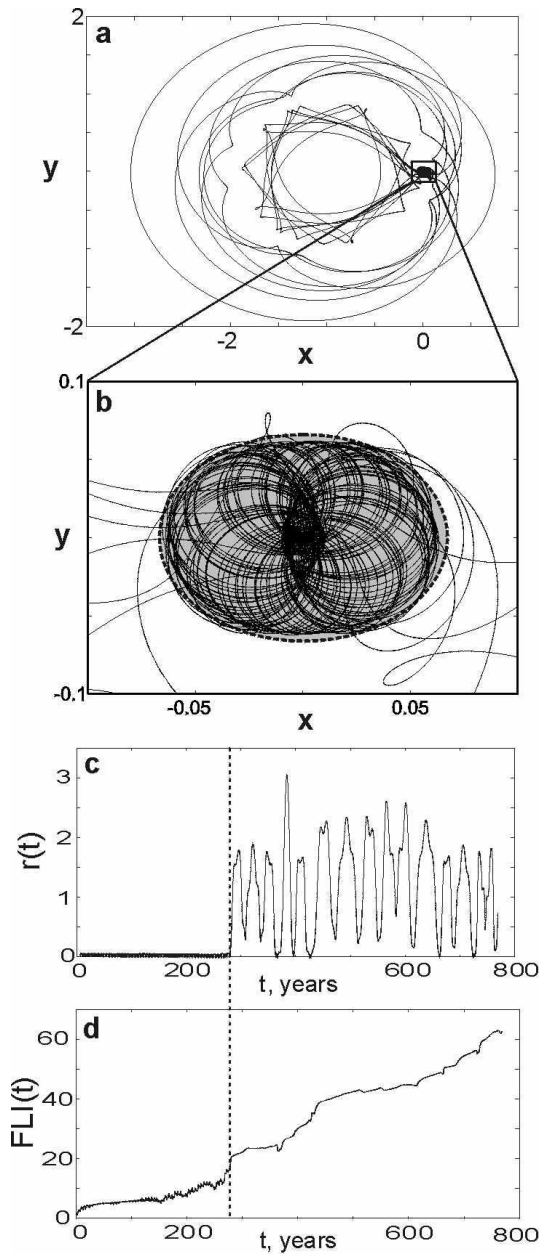}
\caption{\label{escape} A paradigmatic escaping `sticky' trajectory (a) in the planar Sun-Jupiter CRTBP; its multiple returns (b) to the Hill sphere
  (shaded grey, $R_H \simeq 0.068$) after heliocentric excursions;
  its radial distance from the planet $r=\sqrt{x^2+y^2}$ (c) and time profile of
  the Fast Lyapunov Indicator (d). The dotted vertical line indicates
  the first crossing of the Hill sphere (escape).}
\end{center}
\end{figure}

Level curves of the ZVS, similar to a potential energy surface
(PES), serve to limit the motion in the rotating frame and so define
an energetically accessible region that may intersect the Hill
sphere. However, unlike a PES, because the ZVS is defined in a
rotating frame, it is possible for energy maxima to be stable as is,
in fact, the case for $L_4$ and $L_5$ at which points Jupiter's
Trojan asteroids are situated \citep{murray}. The Hill sphere (see
Fig.~\ref{zvs}) roughly occupies the region between the saddle
points $L_1$ and $L_2$ which we term the `capture zone' with its
radius being given by $R_H=a_p/(\mu/3)^{1/3}$ where $a_p$ is the
planet's semimajor axis \citep{murray}. In the case of ERTBP the ZVS
pulsates, which defines periodically time-dependent capture regions
\citep{mako}. The two Lagrange saddle points $L_1$ and $L_2$ act as
gateways between the Hill sphere and heliocentric orbits. A key
finding of \citet{cac2003} is that at energies close to (but above)
the Lagrange points only prograde orbits can enter (or exit) the
capture zone. At higher capture energies the distribution shifts to
include both senses of $h_z$ until, finally, retrograde capture
becomes more likely. The statistics of inclination distributions
will, therefore, be expected to depend strongly on energy, i.e., how
the curves of zero-velocity intersect the Hill sphere.
Fig.~\ref{sos} portrays the structure of phase space in the planar
limit ($z = p_z = 0$) in a series of Poincar\'e surfaces of section
at four energies. At the lowest energy shown in Fig.~\ref{sos}(a)
many of the prograde orbits are chaotic whereas all the retrograde
orbits are regular. Because incoming orbits cannot penetrate the
regular KAM tori, prograde orbits must remain prograde while
retrograde orbits cannot be captured nor can already bound
retrograde orbits escape.  After $L_2$ has opened in
Fig.~\ref{sos}(b) the chaotic `sea` of prograde orbits quickly
`evaporates' except for a `sticky' layer of chaos which clings to
the KAM tori.  With increasing energy this front evolves from
prograde to retrograde motion. Chaotic orbits close to the remaining
tori can become trapped in almost regular orbits for very long
times. In the presence of dissipation these chaotic orbits can be
smoothly switched into the nearby KAM region and almost always
preserve the sign of angular momentum.  Thus, at low (high) energy
permanent capture is almost always into prograde (retrograde)
orbits.

\section{Results and discussion}

In this section we describe the results of our simulations in the planar ERTBP using the FLI and our Monte Carlo simulations in the spatial ERTBP.

\subsection{Planar ERTBP and FLI}

\begin{figure}
\begin{center}
\includegraphics[width=7cm]{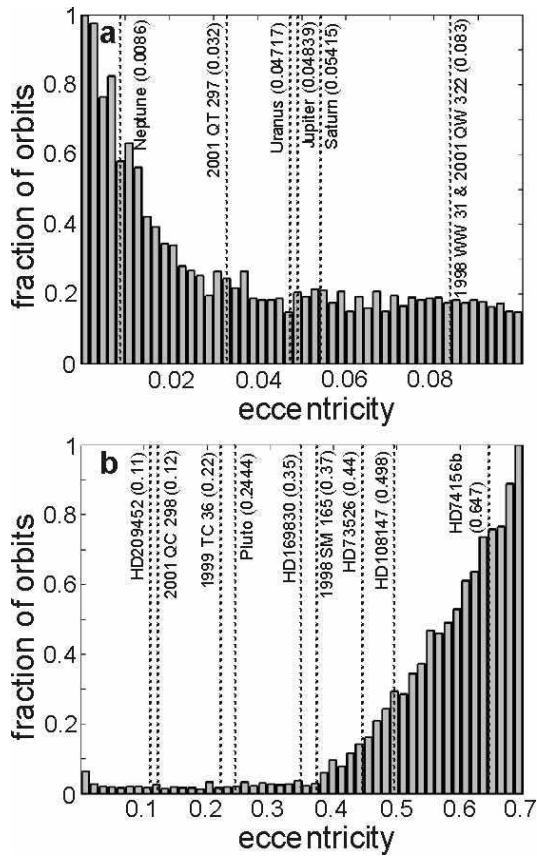}
\caption{\label{fract} Fraction of chaotic orbits within the Hill sphere that have $FLI > 8$ at
the cut-off time $T_{cut}=200$ years in the planar Sun-Jupiter ERTBP as a function of eccentricity. The phase
space was sampled randomly with initial energies, true anomalies and eccentricities taken also at random.
For reference, shown in parentheses are the mean eccentricities of Jupiter, Saturn, Neptune, Uranus, Pluto,
some extrasolar planetary systems (HD 209452, HD 169830, HD 73526, HD 108147, HD 74156b),
and binary trans-Neptunian objects on elliptic orbits
($2001~QC_{297}$,$1998~WW_{31}$, $2001~QW_{322}$, $2001~QC_{298}$, $1999~TC_{36}$, $1998~SM_{165}$).}
\end{center}
\end{figure}

We are primarily interested here in obtaining a qualitative picture of
the volume and structure of phase space occupied by the chaotic layer in the ERTBP
as compared to the CRTBP. For this we find a good diagnostic to be
$FLI(t)$ whose increase for chaotic orbits can usually be detected
before, or no later than, a test particle finally escapes from
the capture zone (Fig.~\ref{escape}).  These measurements, when made
over the capture region (which contains permanently bound regular
trajectories {\it even at energies well above $L_1$ and $L_2$}), provide an estimate of the number of orbits that
could be captured as illustrated in Fig.~\ref{fract}. The number of
chaotic orbits inside the capture zone, detected by computing FLI, decreases with increasing
ellipticity for moderate eccentricities (Fig.~\ref{fract}(a)) signifying that the chaotic layers
get weaker and, therefore, capture is expected to become less probable than it is at $e=0$.

Ideally the problem of separating the fractions of temporarily
trapped and almost immediately escaping trajectories could be better
quantified by partitioning the phase space into disjoint (e.g.
`inner', Hill sphere, and `outer', heliocentric space) regions and
computing the fluxes across the barriers between them. But, given
the current state of phase space transport theory, this has not yet
been shown possible in practice for essentially 3D problems such as
the spatial CRTBP and ERTBP Hamiltonians. Interestingly enough,
despite strong theoretical grounds \citep*{wiggins}, and exhausting
attempts, efforts to describe quantitatively spatial three-body
dynamics by constructing global invariant manifolds
\citep{belbruno}, that would presumably contain all possible
incoming and outgoing chaotic trajectories through $L_1$ and $L_2$
saddle points, have been not quite successful in approaching the 3D
problem so far. Part of the reason is that multiple escapes and
recurrences of high energy trajectories to the Hill sphere (see
example on Fig.~\ref{escape}a,b) make local manifolds near
multidimensional saddles extremely difficult to use as rigorous
surfaces of no return. This is even more pronounced for chaotic
ionization of atomic Rydberg electrons (\citealt*{atom};
\citealt{lee}) which is closely related to RTBP dynamics. Recent
progress in pursuit of manifolds for the spatial three-body problem
is reported by \citealt{marsden, scheeres}; \citealt*{waalkens}.

\begin{figure*}
\begin{center}
 \includegraphics{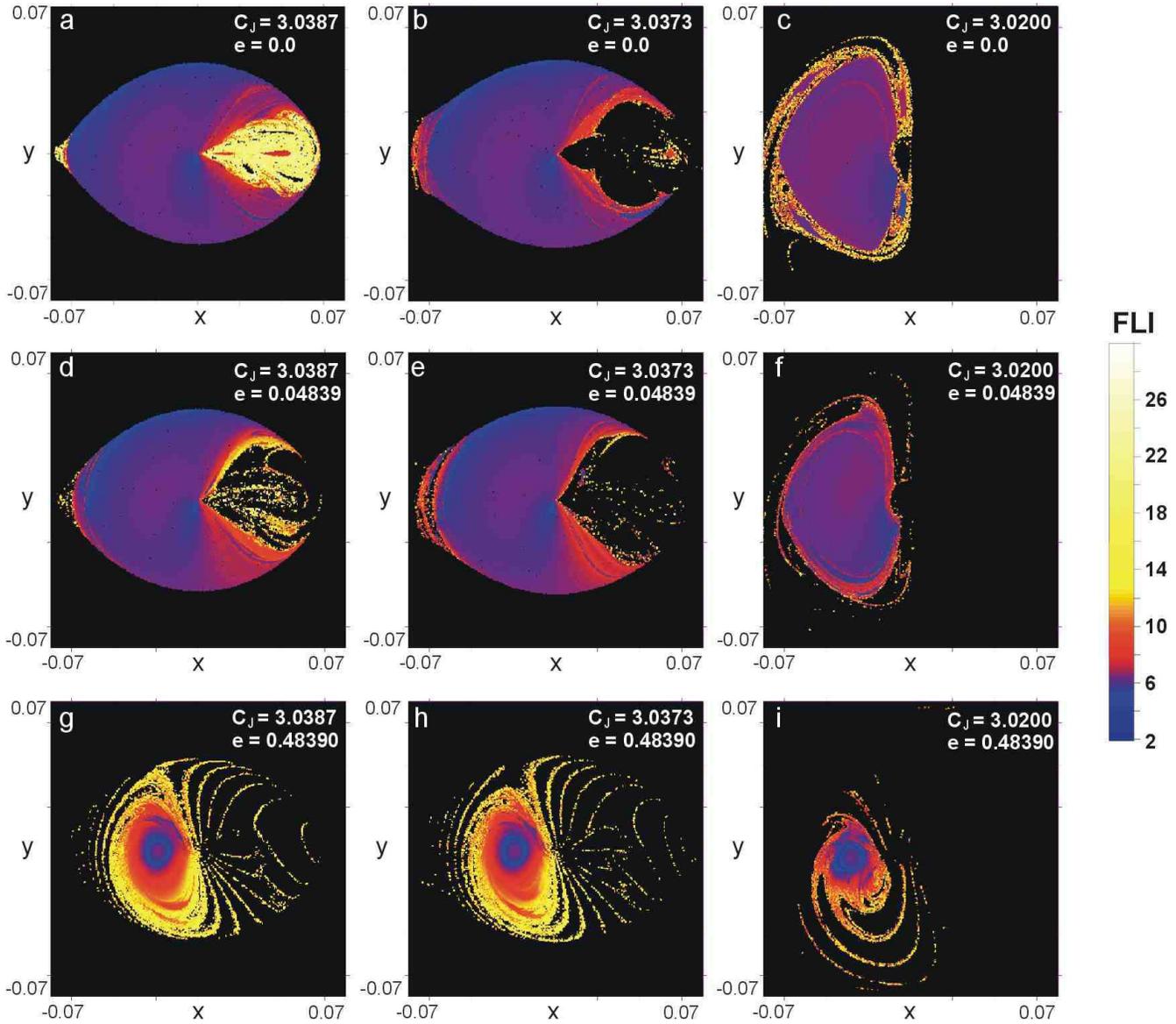}
 \caption{Colour coded Fast Lyapunov Indicator on the grid
 of isoenergetic (at indicated values of the Jacobi constant $C_J$) initial conditions
 taken on the $x-y$ surface of section (${p_x}=0, \dot{y}>0$) for the planar circular (a--c), elliptic Sun-Jupiter (d--f) and and highly elliptic extrasolar (g--i) systems.
 In the elliptic case, initial conditions were computed with initial $f=\pi/2$. FLI were measured at the cut-off time of
 $T_{cut}=200$ years. Very short-lived (scattering) trajectories that did not survive inside the Hill sphere
 for the cut-off time were discarded and are not shown.}
 \label{FLI}
\end{center}
\end{figure*}

To analyse the structure of phase space in ERTBP, we first computed
FLI in the planar (2D) circular case ($e = 0, z = p_z = 0$), where
direct comparison with surfaces of section (in the Hill limit, $\mu
\ll 1$, \citealt{simo, cac2003}) can be made. Figs.~\ref{FLI}(a--c)
confirm that phase space visualisation via computing short-time FLI
(\ref{fli}) works well in the planar CRTBP, reproducing correctly
all the relevant features visible in the SOS shown in
Fig.~\ref{sos}. In particular, the chaotic layer, and its evolution
with increasing energy, can be easily identified by high values of
FLI (shown in yellow). Note also that FLI measurements make sense
even for relatively short time integrations, and are more economical
than constructing the corresponding SOS (compare cut-off times given
in the Fig.~\ref{FLI} and Fig.~\ref{sos} captions). As for the
sensitivity, it is not clear how reliable the numerical distinctions
(red-purple-blue) between the FLI for quasiperiodic and resonant
orbits are, but this is irrelevant for our current purposes.

In the planar limit, initial conditions were generated randomly within
the Hill radius on the surface of section (see Fig.~\ref{FLI} caption)
and assuming that, initially, $f = \pi/2$ in which
case the ERTBP initial conditions reduce to those of CRTBP. This guaranties that all ERTBP initial
conditions are generated with identical initial energies $p_f(0)=-E_e$, and these are, in fact,
true CRTBP energies $E_c$. The setting, thereby, allows for a direct
comparison between the SOS of Fig.~\ref{sos} and results obtained using the
FLI. Due to the dimensionality of the ERTBP, relaxation of the above constraint, e.g. choosing the
initial true anomalies at random, produced FLI maps with no distinct structure.
On the contrary, consistency in initial conditions at fixed $f = \pi/2$
helps track the smooth changes from CRTBP to ERTBP.

As the planet's orbital eccentricity is increased to reach its actual value for Jupiter,
FLI maps reveal a reduction of the density of orbits within the chaotic layers (yellow paterns) visible
on Fig.~\ref{FLI}(d-f) at each of the corresponding energies shown. However, at this moderate
eccentricity, the phase space structures responsible for CAC (`sticky' KAM tori surrounded by the
chaotic layers) generally survive any deviations introduced by ellipticity. This suggests robustness of the
CAC mechanism with respect to actual ellipticities of the giant planets' orbits which lie well
below $e \simeq 0.4$ (see Fig.~\ref{fract}). We further confirm this by direct Monte Carlo simulations of capture probability (Sec.~4.2) which reveal
that CAC indeed survives weak ellipticity in the RTBP. An important implication
for future studies can be drawn by noticing that the heliocentric orbits of Pluto (with Charon)
as well as of some recently discovered binary trans-Neptunian objects
\footnote{\tt http://www.johnstonsarchive.net/astro/asteroidmoons.html}
are quite eccentric (see examples on Fig.~\ref{fract}), but not to the extent characteristic
of extrasolars (although the relative orbit of the binary partners as distinct from the heliocentric orbit of the
combine can be very eccentric).
These binary objects can be viewed in the framework of the Hill approximation too, so the weakly elliptic chaos-assisted mutual capture, probably stabilized by the fast exchange of energy and angular momentum
with a fourth body, may have been involved in early stages of their formation (\citealt*{gold};
\citealt{cac2003}; \citealt{hut}).

The picture at much higher eccentricities evolves towards significant distorsions of the
phase space stuctures as visualised by FLI. In Fig.~\ref{FLI}(g--i) the KAM tori (red--blue
islands) shrink, giving way to regions of scattering and chaotic orbits with short (200 years
in this example) residence times inside the Hill sphere. These short-lived trajectories are the
main contributions to the rapidly increasing number of chaotic orbits observed in Fig.~\ref{fract}
for $e > 0.4$. This does not mean, however, that capture will necessarily be enhanced, because a high density
of strongly chaotic orbits does not correlate with the number of very long-lived trajectories
trapped close to KAM tori.

\begin{figure}
\begin{center}
\includegraphics[width=8cm]{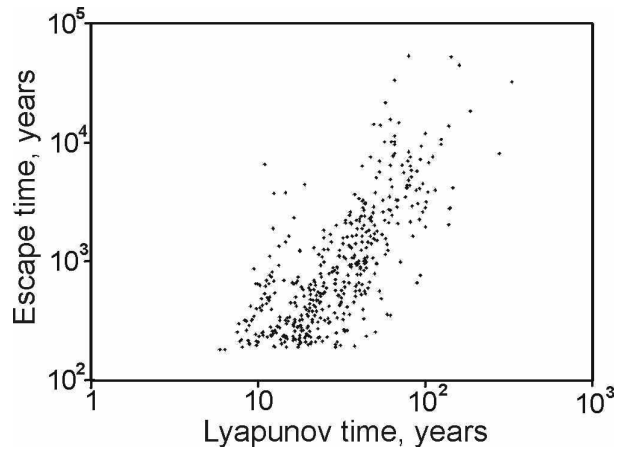}
\caption{\label{L_times} Escape time {\it versus} Lyapunov time for 2D CRTBP chaotic trajectories
with initial conditions chosen randomly inside the Hill sphere. Escape was defined as the
first crossing of the Hill sphere.}
\end{center}
\end{figure}

On the other hand, Lyapunov exponents computed for individual trajectories
cannot serve as a predictor of global stability, i.e. as an indicator of whether an
orbit will stay long enough in a bound region, or if it escapes quickly. This aspect
in computing Lyapunov exponents for escaping (captured)
trajectories concerns the notion of `stable
chaos' \citep{stable}. This term was coined to refer to chaotic motions
that are locally hyperbolic, but demonstrate macroscopic
long-term stability, so that the Lyapunov time $T_L$ (the inverse of
the largest Lyapunov exponent (\ref{lyap})) is substantially less than
the 'event' time $T_E$, e.g. the time for a trajectory to leave a
certain region. However, there are indications of a possible simple
linear correlation between $T_E$ and $T_L$ as shown in \citet*{murison},
which suggests that knowledge of the Lyapunov time could allow one to
predict an 'event' time.  Concerning the residence (escape)
time near the planet, simulations in the planar CRTBP, however, show
that such a correlation with the Lyapunov time does not exist
(Fig.~\ref{L_times}, see also discussion by \citealt{morbidelli} and \citealt{varvoglis}).
Among the randomly chosen escaping chaotic trajectories
originating in the Hill sphere, there are many very long-lived examples that
have diverse, even quite small, Lyapunov times. This means that local
(microscopic) instabilities in the chaotic layer alone cannot explain
the distributions of survival probability.  Rather, as was pointed out
by \citet*{tsiganis}, long-term trapping with short $T_L$ near `sticky'
KAM structures can be attributed to the existence of phase space
quasi-barriers and approximate integrals of motion. In the case of
RTBP, these are the phase space structures around the saddle points at $L_1$ and $L_2$
and the $z$-component of angular momentum with respect to the planet, respectively.
In 3D, the latter is approximately conserved \citep{conto} for long-lived chaotic orbits
which explains the unexpected strong correlation between initial and final
inclinations of captured particles \citep[`inclination memory',][]{cac2003}.

\subsection{Monte Carlo simulations in the spatial ERTBP}

\begin{figure}
\includegraphics[width=8cm]{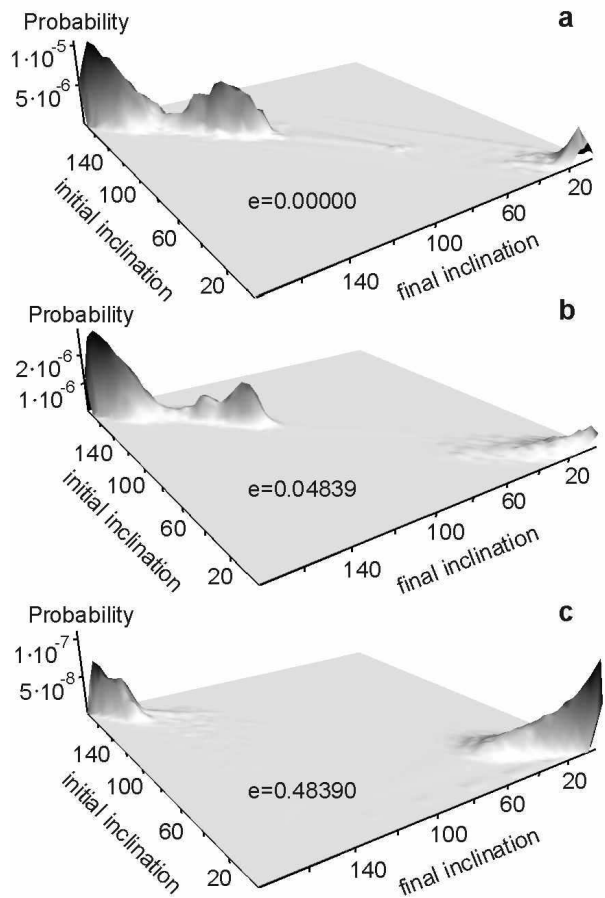}
\caption{\label{prob} Capture probability (the density of survivors normalized by the number of trajectories in the incoming fluxes)
as a function of initial and final inclinations in spatial circular (a), elliptic Sun-Jupiter (b)
and exemplary elliptic extrasolar (c) RTBP systems. The values of eccentricity $e$ are shown.
Chaotic trajectories with random energies ($C_J \in [2.995,{C_J}^{(L_1)}]$), true anomalies
($f \in [0,2\pi]$), velocities and coordinates (on the Hill sphere with the radius $R_H \simeq 0.068$)
were integrated until they either escaped from the sphere or crossed Callisto's orbit
at $r=2.42 \times 10^{-3}$ (Jupiter's orbital semimajor axis $a_p=1$).
Each of the distrubutions was drawn from orbital elements of $N \simeq 20000$ long-lived chaotic
orbits that entered and survived within the Hill sphere for $T_{cut}=20000$ years. Incoming fluxes consisted
of $3 \times 10^7$ (a), $1.6 \times 10^8$ (b), $4.6 \times 10^9$ (c) particles.}
\end{figure}

We simulated capture statistics in the spatial (three-dimensional) ERTBP
by integrating isotropic fluxes of
test particles that bombard the capture zone producing equal
probabilities of initial conditions. Initial conditions were generated as follows: the
particle's position vector was chosen uniformly and randomly on the
surface of the Hill sphere.  Velocities were also chosen uniformly and
randomly in accordance with the value of the CRTBP energy $E_c$.
In turn, the starting energy (Jacobi constant $C_J=-2E_c$) was chosen uniformly random between its minimum
possible value (as defined by the energy of $L_1$ when
$f = \pi/2$) and its highest value (determined empirically
such that above it the capture probability was essentially zero.)
Then, by randomizing the initial true anomaly $f$ we model equal chances
for a test particle to have any phase with respect to the mutual revolution of
the primaries. The trajectories were integrated until one of the following occurred: the particle exited
the Hill sphere; it penetrated a sphere, centred on the planet, of a
given radius (see Fig.~\ref{prob} caption); it survived for a predetermined
cut-off time.

Choosing particles on the Hill sphere, where the motion is chaotic or
scattering, minimizes (although does not eliminate) the risk of
accidentally starting orbits inside impenetrable KAM regions \citep{neto2004}.
Although permanently bound, these orbits could never actually have
been captured because KAM regions cannot be penetrated at all in 2D
and only exponentially slowly in 3D.  It is only in the chaotic
layer between scattering and stability that capture can happen.
An initial swarm of incoming particles produces broad distribution of survivors with different residence times inside the Hill sphere. We monitored its dynamics up to certain cut-off times
(of the order of several thousand years) to find those
long-lived chaotic orbits that may have been vulnerable to capture by a relatively weak dissipative force such as gas drag.
Only these orbits, in the CAC model, could be the precursors to the currently observed distributions of irregular satellites.
In a statistical sense, any short-time flybys are unlikely to have contributed to the primordial population of potential moons. Also, since
the Sun-Jupiter system was chosen as an example, we eliminated test particles that penetrated
the inner region of the Hill sphere occupied by Jupiter's most influencial regular moon Callisto.
The massive regular moons may have acted as a source of strong perturbation removing
some temporarily captured moonlets following prograde orbits. This provides a possible reason for the
observed prograde-poor distribution of jovian irregulars \citep{cac2003}.

Fig.~\ref{prob} shows the results of these simulations. The capture probability density
is plotted on the plane of initial and final inclinations.
Unlike other properties, e.g., energy, inclination can be defined consistently
in both the CRTBP and the ERTBP ($e=0$ and $e > 0$ in isotropically pulsating coordinates). As the distributions in
Fig.\ref{prob} confirm, inclination is approximately conserved during temporary capture in the ERTBP, so that
the RTBP energy/inclination dependent CAC mechanism survives additional perturbations caused
by deviations of the planet's orbit from circularity. This is clearly seen upon comparing
Fig.\ref{prob} (a) and (b). The latter shows the capture probability for the Sun-Jupiter system
with its actual eccentricity being used. We also verified by similar Monte Carlo
runs that the basic energy--inclination correlation discussed in Sec.~3 remains valid in the 3D ERTBP.
We note, however, that since the energy is not a conserved quantity in the elliptic problem,
a better representation of results is given in terms of inclinations.

The most prominent effect observed in the ERTPB compared to the
circular case is the overall decrease of capture probability as
eccentricity increases. Simulations indicate that, given the same
time scale, the actual ellipticity of Jupiter's orbit accounts for
approximately an order of magnitude lower capture probability than
predicted by the circular model, while the relative number of
progrades {\it versus} retrogrades remains unaffected.

Capture by typical extrasolar planets will be expected to become even more
suppressed due to the wildly eccentric nature of these, essentially ERTBP, star--planet systems.
To test this we used an eccentricity $e=0.4839$ which is ten times greater than that of
Jupiter's orbit but quite similar to what has been estimated, for instance, for several already
detected extrasolars (the values of $e$ are given in parentheses): HD 108147 (0.498),
HD 168443b (0.53), HD 82943c (0.54), HD 142415 (0.5), HD 4203 (0.46), HD 210277 (0.46), HD 147513 (0.52),
HD 190228 (0.5), HD 50554 (0.5), HD 33636 (0.53). For consistency, we left unchanged all of the other
parameters and environmental factors (including Callisto whose hypothetical extrasolar analogs may well
exist around exo-Jupiters) exactly as they were used in the simulations for Sun-Jupiter. Fig.\ref{prob}(c) indicates that,
due to the ellipticity of the orbits and assuming similar time scales and other
conditions, capture by extrasolar planets may be as much as aproximately ten times less efficient than
it could be by the giant planets of our system.

The relatively low capture probability, predicted in ERTBP, adds to other destructive mechanisms
possibly operating on as yet undiscovered irregular satellites of extrasolar planets. The loss
of satellites through Yarkovsky \citep{death} or tidal \citep{barnes} effects may also diminish
the possibility for highly elliptic extrasolar captors to marshal large
populations of irregular moons. On the other hand, exo-planets with masses from one-half
to four times Jupiter's mass on low eccentricity orbits ($e < 0.02$) may be considered candidates
for having irregular satellites. These may include, e.g., already known extrasolars
such as HD 179949, 55 Cnc b, HD 169830c, HD 187123, Tau Boo, HD 75289, 51 Peg, Ups And b, HD 195019.

\section{Conclusions}

The complexity of many-body systems evidently goes far beyond that
of the simplest non-trivial case, the three-body problem. However,
when considered on a hierarchical level, interactions between
gravitional centers can often be decomposed to relatively low
degree-of-freedom subsystems for which analysis from a dynamical
systems point of view becomes possible. In the problem described
here, the existence of 'sticky' volumes of phase space where regular
regions are surrounded by chaotic layers explains why several bodies
may become temporarily (but for rather long time periods) trapped by
mutual chaos-assisted capture. These quasi-stable configurations can
subsequently be stabilised (or destroyed) on much longer time scales
by various forces (dissipative or not). Here we have considered the
effect introduced by a slow (compared to time-scales of motion in
the chaotic layer) periodic parametric time dependence in the
three-body problem. Our simulations show that chaos-assisted capture
applies and is, in fact, only slightly perturbed in cases when the
primaries move on an elliptic orbit. It is expected to be of
significant interest (e.g. in applications to essentially
many-particle systems such as star clusters and the asteroid belt)
to consider chaos-assisted capture at the next hierarchical level,
i.e. when the primary two-body configuration is not restricted to a
regular orbit, which may be the case in three-body encounters of
comparably sized objects.

\section*{acknowledgments}
  This work was supported by grants from the US National Science
  Foundation through grant 0202185 and the Petroleum Research Fund
  administered by the American Chemical Society. All opinions
  expressed in this article are those of the authors and do not
  necessarily reflect those of the National Science Foundation. S.~A.~A.
  also acknowledges support from Forschungszentrum J\"ulich, where
  part of this work was done.


\begin{thebibliography}{7}

\bibitem[\protect\citeauthoryear{Aarseth}{2003}]{Aarseth}
Aarseth S.~J., 2003, Gravitational N-body Simulations: Tools and Algorithms. Cambridge Univ. Press, Cambridge

\bibitem[\protect\citeauthoryear{Astakhov et al.}{2003}]{cac2003}
Astakhov S.~A., Burbanks A.~D., Wiggins S., Farrelly D., 2003, Nature, 423, 264

\bibitem[\protect\citeauthoryear{Barnes \& O'Brien}{2002}]{barnes}
Barnes J.~W., O'Brien D.~P., 2002, ApJ, 575, 1087

\bibitem[\protect\citeauthoryear{Beaug\'e \& Michtchenko}{2003}]{beauge}
Beaug\'e C., Michtchenko T.~A., 2003, MNRAS, 341, 760.

\bibitem[\protect\citeauthoryear{Benest}{2003}]{benest}
Benest D., 2003, A\&A, 400, 1103

\bibitem[\protect\citeauthoryear{Belbruno}{2004}]{belbruno}
Belbruno E., 2004, Capture Dynamics and
Chaotic Motions in Celestial Mechanics. Princeton Univ. Press, Princeton and Oxford

\bibitem[\protect\citeauthoryear{Brunello, Uzer \& Farrelly}{Brunello et al.}{1997}]{atom}
Brunello A.~F., Uzer T., Farrelly D., 1997, Phys. Rev. A, 55, 3730

\bibitem[\protect\citeauthoryear{Burns \& \'Cuk}{2002}]{death}
Burns J.~A., \'Cuk M., 2002, BAAS, 34, DPS 34th Meeting, abstr. No. 42.01

\bibitem[\protect\citeauthoryear{Carruba et al.}{2002}]{carruba2002}
Carruba V., Burns J.~A., Nicholson P.~D., Gladman B.~J., 2002, Icarus, 158, 434

\bibitem[\protect\citeauthoryear{Chiang, Fischer \& Thommes}{Chiang et al.}{2002}]{fischer}
Chiang E.~I., Fischer D., Thommes E., 2002, ApJ, 564, L105

\bibitem[\protect\citeauthoryear{Chodas \& Yeomans}{1996}]{chodas}
Chodas P.~W., Yeomans D.~K., 1996, in Knoll K.~S., Weaver H.~A.,
Feldman P.~D., ed., The collision of Comet
 Shoemaker-Levy 9 and Jupiter. Cambridge Univ. Press, Cambridge

\bibitem[\protect\citeauthoryear{Cincotta, Giordano \& Sim\'o}{Cincotta et al.}{2003}]{simo2003}
Cincotta P.~M., Giordano C.~M., Sim\'o C., 2003, Physica D, 182, 157

\bibitem[\protect\citeauthoryear{Contopolous}{1965}]{conto}
Contopoulos G., 1965, ApJ, 142, 802

\bibitem[\protect\citeauthoryear{\'Cuk \& Burns}{2004}]{burns2004}
\'Cuk M., Burns J.~A., 2004, Icarus, 167, 369

\bibitem[\protect\citeauthoryear{Froeschl\'e, Guzzo \& Lega}{Froeschl\'e et al.}{2000}]{fro2000}
Froeschl\'e C., Guzzo M., Lega E., 2000, Science, 289, 2108

\bibitem[\protect\citeauthoryear{Funato et al.}{2004}]{hut}
Funato Y., Makino J., Hut P., Kokubo E., Kinoshita D., 2004, Nature, 427, 518

\bibitem[\protect\citeauthoryear{Gladman et al.}{2001}]{gladman2001}
Gladman B. et al., 2001, Nature, 412, 163

\bibitem[\protect\citeauthoryear{Goldreich \& Sari}{2003}]{goldreich}
Goldreich P., Sari R., 2003, ApJ, 585, 1024

\bibitem[\protect\citeauthoryear{Goldreich, Lithwick \& Sari}{Goldreich et al.}{2002}]{gold}
Goldreich P., Lithwick Y., Sari R., 2002, Nature, 420, 643

\bibitem[\protect\citeauthoryear{G\'omez et al.}{2003}]{marsden}
G\'omez G., Koon W.~S., Lo M.~W., Marsden J.~E., Masdemont J., Ross
S.~D., 2004, Nonlinearity, 17, 1571

\bibitem[\protect\citeauthoryear{Grav \& Holman}{2004}]{grav2004}
Grav T., Holman M.~J., 2004, ApJ, 605, L141

\bibitem[\protect\citeauthoryear{Kavelaars et al.}{2004}]{kavel}
Kavelaars J.~J. et al., 2004, Icarus,
169, 474

\bibitem[\protect\citeauthoryear{Lecar, Franklin \& Murison}{Lecar et al.}{1992}]{murison}
Lecar M., Franklin F., Murison M., 1992, AJ, 104, 1230

\bibitem[\protect\citeauthoryear{Lee et al.}{2000}]{lee}
Lee E., Brunello A.~F., Cerjan C., Uzer T., Farrelly D., 2000, in Yeazell J., Uzer T., ed., The Physics and
Chemistry of Wave Packets, Wiley, NY, p. 95

\bibitem[\protect\citeauthoryear{Lichtenberg \& Lieberman}{1992}]{ll}
Lichtenberg A.~J., Lieberman  M.~A., 1992, Regular and Chaotic Dynamics, 2nd edn.
Springer-Verlag, NY

\bibitem[\protect\citeauthoryear{Llibre \& Pi\~nol}{1990}]{llibre}
Llibre J., Pi\~nol J., 1990, Celest. Mech. Dynam. Astronom., 48, 319

\bibitem[\protect\citeauthoryear{Lorenz}{1965}]{lorenz}
Lorenz E.~N., 1965, Tellus, 17, 321

\bibitem[\protect\citeauthoryear{Mak\'o \& Szenkovits}{2004}]{mako}
Mak\'o Z., Szenkovits F., 2004, Celest. Mech. Dynam. Astronom., in
press


\bibitem[\protect\citeauthoryear{Marcy \& Butler}{2000}]{marcy}
Marcy G.~W., Butler R.~P., 2000, PASP, 112, 137

\bibitem[\protect\citeauthoryear{Marzari \& Weidenschilling}{2002}]{marzari}
Marzari F., Weidenschilling S.~J., 2002, Icarus, 156, 570

\bibitem[\protect\citeauthoryear{Michtchenko \& Malhotra}{2004}]{michtchenko}
Michtchenko T.~A., Malhotra R., 2004, Icarus, 168, 237

\bibitem[\protect\citeauthoryear{Milani \& Nobili}{1992}]{stable}
Milani A., Nobili A.~M., 1992, Nature, 357, 569

\bibitem[\protect\citeauthoryear{Morbidelli \& Froeschl\'e}{1996}]{morbidelli}
Morbidelli A., Froeschl\'e C., 1996, Celest. Mech. Dynam. Astronom., 63, 227

\bibitem[\protect\citeauthoryear{Murray \& Dermott}{1999}]{murray}
Murray C.~D., Dermott S.~F., 1999, Solar System Dynamics. Cambridge Univ. Press, Cambridge

\bibitem[\protect\citeauthoryear{Nagler}{2004}]{nagler}
Nagler J., 2004, Phys. Rev. E, 69, 066218


\bibitem[\protect\citeauthoryear{Nesvorn\'y et al.}{2003}]{nesvorny2003}
Nesvorn\'y D., Alvarellos J.~L.~A., Dones L., Levison H.~F., 2003,
AJ, 126, 398

\bibitem[\protect\citeauthoryear{Nesvorn\'y, Beaug\'e \& Dones}{Nesvorn\'y et al.}{2004}]{nesvorny2004}
Nesvorn\'y D., Beaug\'e C., Dones L., 2004, AJ, 127, 1768

\bibitem[\protect\citeauthoryear{Neto, Winter \& Yokoyama}{Neto et al.}{2004}]{neto2004}
Neto E.~V., Winter O.~C., Yokoyama T., 2004,  A\&A, 414, 727

\bibitem[\protect\citeauthoryear{Pilat-Lohinger, Funk \& Dvorak}{Pilat-Lohinger et al.}{2003}]{dvorak}
Pilat-Lohinger E., Funk B., Dvorak R., 2003, A\&A, 400, 1085

\bibitem[\protect\citeauthoryear{Press et al.}{1999}]{NR}
Press W.~H., Teukolsky S.~A., Vetterling W.~T., Flannery B.~P., 1999,
Numerical Recipes in C, 2nd edn.  Cambridge Univ. Press, Cambridge

\bibitem[\protect\citeauthoryear{S\'andor et al.}{2001}]{sandor}
S\'andor Z., Balla R., T\'eger F., \'Erdi B., 2001, Celest. Mech. Dynam. Astronom., 79, 29

\bibitem[\protect\citeauthoryear{S\'andor et al.}{2004}]{sandor2004}
S\'andor Z., \'Erdi B., Sz\'ell A., Funk B., 2004, Celest. Mech.
Dynam. Astronom., in press


\bibitem[\protect\citeauthoryear{Schneider}{1999}]{schneider}
Schneider J., 1999, CR Acad. Sci. II B, 327, 621

\bibitem[\protect\citeauthoryear{Sheppard \& Jewitt}{2003}]{jewitt2003}
Sheppard S., Jewitt D., 2003, Nature, 423, 261

\bibitem[\protect\citeauthoryear{Sim\'o \& Stuchi}{2000}]{simo}
Sim\'o C., Stuchi T.~J., 2000, Physica D, 140, 1

\bibitem[\protect\citeauthoryear{Stiefel \& Scheifele}{1971}]{KS}
Stiefel E.~L., Scheifele G., 1971, Linear and Regular Celestial Mechanics: Perturbed Two-Body Motion,
Numerical Methods, Canonical Theory. Springer-Verlag, New York

\bibitem[\protect\citeauthoryear{Szebehely}{1967}]{sze}
Szebehely V., 1967, Theory of Orbits: the Restricted Problem of Three Bodies.
Acad. Press, NY and London

\bibitem[\protect\citeauthoryear{Tancredi, S\'anchez \& Roig}{Tancredi et al.}{2001}]{tancredi}
Tancredi G., S\'anchez A., Roig F., 2001, AJ, 121, 1171

\bibitem[\protect\citeauthoryear{Tsiganis, Anastasiadis \& Varvoglis}{Tsiganis et al.}{2000}]{tsiganis}
Tsiganis K., Anastasiadis A., Varvoglis H., 2000, Chaos, solitons and fractals, 11, 2281

\bibitem[\protect\citeauthoryear{Tremaine \& Zakamska}{2003}]{tremaine}
Tremaine S., Zakamska N.~L., 2003, preprint (astro-ph/0312045)

\bibitem[\protect\citeauthoryear{Varvoglis \& Anastasiadis}{1996}]{varvoglis}
Varvoglis H., Anastasiadis A., 1996, AJ, 111, 1718

\bibitem[\protect\citeauthoryear{Villac \& Scheeres}{2004}]{scheeres}
Villac B.~F., Scheeres D.~J., 2004, A simple
    algorithm to compute hyperbolic invariant manifolds near $L_1$ and
    $L_2$, 14th AAS/AIAA Space Flight Mechanics Meeting, February
  2004, Maui, Hawaii

\bibitem[\protect\citeauthoryear{Waalkens, Burbanks \& Wiggins}{Waalkens et al.}{2004}]{waalkens}
Waalkens H., Burbanks A., Wiggins S., 2004, J. Phys. A: Math. Gen,
37, L257

\bibitem[\protect\citeauthoryear{Wiggins, Haller \& Mezic}{Wiggins}{1994}]{wiggins}
Wiggins S., Haller G., Mezic I., 1994,
Normally Hyperbolic Invariant Manifolds in Dynamical Systems
(Applied Mathematical Sciences, Vol 105). Springer-Verlag, NY

\bibitem[\protect\citeauthoryear{Williams}{2003}]{williams}
Williams D.~M., 2003, BAAS, 35, DPS 35th Meeting, abstr. No. 27.10

\bibitem[\protect\citeauthoryear{Williams, Kasting \& Wade}{Williams et al.}{1997}]{habitable}
Williams D.~M., Kasting J.~F., Wade R.~A., 1997, Nature, 385, 234

\bibitem[\protect\citeauthoryear{Yokoyama et al.}{2003}]{yokoyama2003}
Yokoyama T., Santos M.~T., Cardin G., Winter O.~C., 2003, A\&A, 401, 763



\end{thebibliography}
\end{document}